\documentclass[12pt]{article}
\usepackage{amssymb, latexsym, amsmath, hyperref}
\usepackage{epsfig}
\usepackage{graphicx}
\usepackage{amsfonts}
\usepackage{mathrsfs}
\usepackage{multirow}
\usepackage[dvips]{color}

\newcommand{\R}{\mathbb{R}}

\newcommand{\N}{\mathbb{N}}

\newcommand{\fD}{\mathfrak{D}}

\newcommand{\cD}{\mathcal{D}}
\newcommand{\cH}{\mathcal{H}}

\newcommand{\cP}{\mathcal{P}}

\newcommand{\cS}{\mathcal{S}}
\newcommand{\cT}{\mathcal{T}}

\newcommand{\be}{\begin{equation}}
\newcommand{\ee}{\end{equation}}
\newcommand{\bea}{\begin{eqnarray}}
\newcommand{\eea}{\end{eqnarray}}

\newcommand{\kt}{\rangle}
\newcommand{\br}{\langle}

\newcommand{\ed}{\end{document}}
\newcommand{\bbr}{\br\!\br}
\newcommand{\kkt}{\kt\!\kt}

\newcommand{\np}{\newpage}

\newcommand{\bi}{\begin{itemize}}
\newcommand{\ei}{\end{itemize}}

\newcommand{\bce}{\begin{center}}
\newcommand{\ece}{\end{center}}

\newcommand{\sD}{\mathscr{D}}

\newcommand{\sH}{\mathscr{H}}

\newcommand{\sR}{\mathscr{R}}

\newcommand{\etap}{{\eta_{_+}}}

\begin{document}

\title{Pseudo-Hermitian Quantum Mechanics with\\ Unbounded Metric Operators}

\author{Ali~Mostafazadeh\thanks{E-mail address:
amostafazadeh@ku.edu.tr, Phone: +90 212 338 1462, Fax: +90 212 338
1559}
\\
Department of Mathematics, Ko\c{c} University,\\
34450 Sar{\i}yer, Istanbul, Turkey}

\date{ }
\maketitle

\begin{abstract}
We extend the formulation of pseudo-Hermitian quantum mechanics to
$\etap$-pseudo-Hermitian Hamiltonian operators $H$ with an unbounded metric operator $\etap$. In particular, we give the details of the construction of the physical Hilbert space, observables, and equivalent Hermitian Hamiltonian for the case that $H$ has a real and discrete spectrum and its eigenvectors belong to the domain of $\etap$ and consequently $\sqrt\etap$.
\vspace{2mm}

\vspace{2mm}

\end{abstract}

Pseudo-Hermitian quantum mechanics is a representation of the conventional quantum mechanics that allows for describing unitary quantum systems using non-Hermitian Hamiltonian operators $H$ whose Hermiticity can be restored by an appropriate change of the inner product \cite{review}.\footnote{Throughout this article we follow von Neumann's terminology of using the term ``Hermitian operator'' to mean ``self-adjoint operator,'' \cite[p 96]{von}.} This theory has emerged \cite{p123,ph,jpa-2003,jpa-2004} as a natural framework for examining the prospects of employing non-Hermitian $\cP\cT$-symmetric Hamiltonians \cite{PT}, such as $H=p^2+ix^3$, in quantum mechanics. Since the publication of \cite{p123} many authors have studied particular examples and various aspects of pseudo-Hermitian operators, and a series of annual international conferences, entitled: ``Pseudo-Hermitian Hamiltonians in Quantum Physics,'' were held  \cite{conf}. These have led to a rapid progress towards solving the basic problems of the subject such as developing methods of constructing inner products, determining the observables of the theory, understanding the role and importance of the antilinear symmetries such as $\cP\cT$-symmetry, and the exploration of the classical limit of pseudo-Hermitian quantum systems. An important technical problem that has resisted a satisfactory resolution is the difficulty associated with the emergence of unbounded metric operators that define the inner product of the physical Hilbert space and the observables of the theory. Indeed for the majority of the toy models studied in the context of pseudo-Hermitian quantum mechanics the metric operator turns out to be  unbounded, while the structure of the theory is developed and fully understood for systems involving bounded metric operators (with a bounded inverse) \cite{review}.

The importance of the complications caused by unbounded metric operators was initially noted in \cite{kresh} where the authors considered a quasi-Hermitian Hamiltonian operator \cite{quasi} with a discrete spectrum and proposed to construct the physical Hilbert space of the system by identifying the inner product with the one making the eigenvectors $\psi_n$ of $H$ orthonormal. This defines an inner-product space on the linear span\footnote{The linear span of $\psi_n$ is the set of all finite linear combinations of $\psi_n$.} of $\psi_n$ which can then be (Cauchy-) completed into a Hilbert space \cite[p 7]{reed-simon}. The main difficulty with this procedure is that it does not give an explicit formula for the inner product of the physical Hilbert space in terms of the information provided by the Hamiltonian operator. Recall that bounded metric operators $\etap$ admit a spectral representation involving a set of eigenvectors of $H^\dagger$ that form a Riesz basis \cite{review}, and the inner product is defined directly in terms of $\etap$. The proposal of Ref.~\cite{kresh} lacks a similar prescription for computing the inner product.

An alternative proposal for dealing with unbounded metric operators is the one given in \cite{jpa-2004,review}. It is based on the view that in pseudo-Hermitian quantum mechanics the (reference) Hilbert space $\sH$, that is used to define the Hamiltonian and the metric operator, does not have a physical significance. This follows from the fact that all the states that can be prepared belong to the domain of the observables one can measure. It is well-known that observables are generally represented by densely-defined linear operators. This in turn means that the physical states are represented by vectors belonging to the intersection $\sD$ of a bunch of dense subsets of $\sH$. In particular, the physical aspects of a given system is fully determined by a proper dense subset $\sD$ of $\sH$ and an inner product that defines the expectation value of the observables. The main message of pseudo-Hermitian quantum mechanics is that this inner product does not need to be the one $\sD$ inherits from $\sH$. The choice of a different inner product on $\sD$ turns it into an inner-product space that can be completed to a Hilbert space $\sH'$. In general $\sH$ and $\sH'$ are different as sets and in particular as topological vector spaces, but both of them include $\sD$ as a dense subset. A simple consequence of this fact is that a metric operator that is defined on $\sD$ may correspond to an unbounded operator with respect to the inner product of $\sH$ while it is a bounded operator with respect to the inner product of $\sH'$. The resolution of the issue of unbounded metric operators that is suggested in \cite{jpa-2004,review} rests on the idea of using $\sH'$ in place of $\sH$ as the reference Hilbert space. In other words, it asserts that because the reference Hilbert space is an axillary mathematical construct, one can choose it in such a way that $\sD$ is a dense subset of the reference Hilbert space and the metric operator of interest acts as a bounded operator in it. This proposal is also difficult to implement in practice, because it does not provide means for an explicit construction of a reference Hilbert space with these properties.

The purpose of the present article is to outline a resolution of the problem of unbounded metric operators that gives an explicit construction for the physical Hilbert space and the observables of the system in terms of an unbounded metric operator.

We first list the basic assumptions upon which our proposal rests:
\begin{enumerate}
\item We fix an infinite-dimensional reference Hilbert space $\sH$ in which the Hamiltonian and other linear operators of interest act as densely-defined closed linear operators. We use the symbol $\br\cdot|\cdot\kt$ to denote the inner product of $\sH$.

\item We consider Hamiltonian operators $H:\sH\to\sH$ with a real and discrete spectrum, so that the linear span of its eigenvectors $\psi_n$, that we denote by $\cS$, is an infinite-dimensional vector subspace of $\sH$.

\item We assume the existence of an unbounded positive-definite operator\footnote{A positive-definite operator, $\pi:\sH\to\sH$, is a self-adjoint operator such that for every nonzero element $\xi$ of its domain the real number $\br\xi|\pi\xi\kt$ is strictly positive.} $\etap:\sH\to\sH$  such that $H$ is $\etap$-pseudo-Hermitian \cite{p123}, i.e., $H$ and its adjoint\footnote{We do not identify linear operators with their matrix representations in some basis, and $H^\dagger$ does not mean complex conjugate of transpose of a matrix. We use the standard mathematical definition of the adjoint of a linear operator \cite[p 252]{reed-simon}. Namely, we let $\cD$ denote the domain of $H$ (which is supposed to be a dense subset of $\sH$) and $\cD':=\{\phi\in\sH|\forall\psi\in\cD,\exists\xi\in\sH, \br\phi|H\psi\kt= \br\xi|\psi\kt\}$. Then $H^\dagger:\sH\to\sH$ is the linear operator with domain $\cD'$ that satisfies the condition: $\forall\psi\in\cD$ and $\forall\phi\in\cD'$, $\br\phi|H\psi\kt=\br H^\dagger\phi|\psi\kt$. We say that $H$ is Hermitian or self-adjoint if $\cD'=\cD$ and for all $\psi,\phi\in\cD$, $\br\phi|H\psi\kt=\br H\phi|\psi\kt$. We say that $H$ is a symmetric operator, if the latter condition holds but $\cD\subseteq\cD'$. In general $\cD'$ may not coincide with $\cD$. Therefore, not every symmetric operator is Hermitian.} $H^\dagger$ fulfil the condition:
        \be
        H^\dagger\etap=\etap H.
        \label{ph}
        \ee
In particular, $H^\dagger\etap$ and $\etap H$ have the same domain.\footnote{This means that for all $\psi\in\sH$ the statement: ``$\psi\in\cD$~{\rm and}~$H\psi\in{\rm Dom}(\etap)$'' is equivalent to ``$\psi\in {\rm Dom}(\etap)$~and~$\etap\psi\in\cD'$.''} An operator $\etap$ with the above properties is called an unbounded metric operator. In view of the positive-definiteness of $\etap$, (\ref{ph}) implies that $H$ is a quasi-Hermitian operator \cite{quasi-old}.\footnote{In \cite{quasi-old} and other rigorous studies of the subject, quasi-Hermiticity is introduced using Eq.~(\ref{ph}) except for the fact that $\etap$ is taken to be a bounded (continuous) linear operator. We use the term quasi-Hermitian in the more general sense where $\etap$ is allowed to be unbounded.}

\item Because $\etap$ is a positive-definite operator, it has a unique positive square root \linebreak $\rho:\sH\to\sH$ that is also a positive-definite operator with a positive-definite inverse $\rho^{-1}$,  \cite[p 281]{kato}. We require that the eigenvectors $\psi_n$ of $H$ belong to the domain of  $\etap$ and consequently $\rho$. This implies that $\rho(\cS)$ is an infinite-dimensional vector subspace of $\sH$.\footnote{One can slightly relax the conditions on $\etap$ by requiring that it is the square of a given invertible (one-to-one) symmetric operator $\rho$ with $\cS$ contained in the domain of $\etap$.}
    \end{enumerate}

Similarly to the proposal of Ref.~\cite{kresh}, we promote $\cS$ to an appropriate inner-product space in which the restriction of $H$ to $\cS$ acts as a Hermitian operator. We do this by endowing $\cS$ with the inner product $\bbr\cdot,\cdot\kkt$ that is defined by the metric operator $\etap$ according to
    \be
    \bbr\phi,\psi\kkt:=\br\phi|\etap\psi\kt=\br\rho\phi|\rho\psi\kt.
    \label{inn-pro}
    \ee
 Here $\phi$ and $\psi$ are arbitrary elements of $\cS$, i.e., they are finite linear combinations of $\psi_n$, and we have used the fact that $\etap=\rho^2$ and $\rho$ is a Hermitian operator. The right-hand side of (\ref{inn-pro}) is finite, because according to Assumption~4, $\psi$ and $\phi$ belong to the domain of $\etap$ and consequently $\rho$. The inner-product space $(\cS,\bbr\cdot,\cdot\kkt)$ obtained in this way can be completed to a Hilbert space that we denote by $\sH_{\etap}$. This is the physical Hilbert space of the pseudo-Hermitian quantum system that we wish to formulate.

We can consider the restriction of $H$ onto $\cS$ and view it as an operator acting in $\sH_\etap$. Then the domain and range of $H$ coincides with $\cS$, and in light of (\ref{ph}), for every pair of elements of $\cS$,  say $\phi$ and $\psi$,
    \bea
    \bbr\phi,H\psi\kkt&=&\br\phi|\etap H\psi\kt=\br\phi|H^\dagger\etap\psi\kt=
    \br H\phi|\etap\psi\kt=\bbr H\phi,\psi\kkt.
    \label{Her}
    \eea
This shows that $H:\cH_{\etap}\to\cH_{\etap}$ is a densely-defined symmetric operator \cite[p 255]{reed-simon}. It has also the appealing property of possessing a complete set of eigenvectors. To see this we recall that because $\cS$ is a dense subset of $\cH_{\etap}$, there is a complete set of eigenvectors of $H$ in $\cH_{\etap}$. Moreover because $H:\cH_{\etap}\to\cH_{\etap}$ is symmetric, the eigenvectors belonging to different eigenspaces are orthogonal. We can perform the Gram-Schmidt process on the eigenvectors belonging to eigenspaces of $H$ to construct orthonormal bases for each eigenspace. The union of these is a complete orthonormal set of eigenvectors of $H$. This is an orthonormal basis of $\cH_{\etap}$. If we label the eigenvectors belonging to such a basis by $\psi_n$ with $n\in\N:=\{0,1,2,\cdots\}$, and the corresponding eigenvalues by $E_n$, we have
    \be
    \bbr\psi_m,\psi_n\kkt=\delta_{mn},~~~~~
    \sum_{i=0}^\infty|\psi_i\kkt\bbr\psi_i|=I,~~~~~
    H\psi_n=E_n\psi_n,
    \label{eg-va}
    \ee
where $m,n\in\N$ are arbitrary, $|\psi_i\kkt\bbr\phi_i|$ stands for the projection operator $\Lambda_i:\cH_{\etap}\to\cH_{\etap}$ defined by:
    \[\forall\chi\in\cH_{\etap},~~\Lambda_i\chi:=\bbr\psi_i,\chi\kkt\psi_i,\]
and $I$ denotes the identity operator acting on $\cH_{\etap}$. Furthermore, because $H$ is a symmetric operator, $E_n$ are necessarily real.

In general, in order to define a quantum system, one needs a Hilbert space that determines the kinematical aspects of the system and a Hamiltonian operator that specifies its dynamics. The latter is required to be a self-adjoint (Hermitian) operator acting in the Hilbert space so that its expectation value in every state is a real number. In the above construction we showed that $H$ acts as a symmetric operator in the Hilbert space $\sH_{_\etap}$. But not every symmetric operator is self-adjoint. Indeed the operator $H:\sH_{_\etap}\to\sH_{_\etap}$ turns out not to be self-adjoint, as its adjoint has a larger domain than that of $H$, \cite[pp 94-95]{exner}. Therefore, we cannot identify it with the Hamiltonian operator for a unitary quantum system. What we can do is to use an appropriate self-adjoint extension of $H$ for this purpose. This is actually very easy and natural to construct.

Recall that because $\{\psi_n|n\in\N\}$ is an orthonormal basis of $\sH_{_\etap}$, every element of $\sH_{_\etap}$ has the form $\sum_{n=0}^\infty a_n\psi_n$, where $\{a_n\}$ is a square-summable sequence of complex numbers, i.e., $\sum_{n=0}^\infty|a_n|^2<\infty$. Now, let $\fD$ be the subset of $\sH_{_\etap}$ consisting of the elements $\sum_{n=0}^\infty a_n\psi_n$ that satisfy the condition: $\sum_{n=0}^\infty E_n^2|a_n|^2<\infty$, i.e.,
    \[\fD:=\left\{\sum_{n=0}^\infty a_n\psi_n~\left|~\sum_{n=0}^\infty E_n^2|a_n|^2<\infty\right.\right\}.\]
Clearly $\cS \varsubsetneq \fD$. Therefore $\fD$ is a dense subset of $\sH_{_\etap}$. Now, we define $\hat H:\sH_{_\etap}\to\sH_{_\etap}$ as the operator that has $\fD$ as its domain and satisfies:
    \be
    \hat H\left(\sum_{n=0}^\infty a_n\psi_n\right):=
    \sum_{n=0}^\infty E_na_n\psi_n.
    \label{hat=}
    \ee
It is clear that $H$ is the restriction of $\hat H$ to $\cS$. Furthermore, it is not difficult to show that $\hat H$ is a self-adjoint operator \cite[p 94]{exner}. Therefore, $\hat H$ is a self-adjoint extension of $H$, and the pair $(\sH_{_\etap},\hat H)$ defines a unitary quantum system. Again the physical condition that the expectation values of observables must be real numbers demands that we identify the observables of this system with the self-adjoint operators acting in $\sH_{_\etap}$ , \cite{review}.

The self-adjoint operator $\hat H$ is actually the closure of $H$. Therefore $H$ is essentially self-adjoint, and $\hat H$ is its unique self-adjoint extension, \cite[p 96]{exner}. This shows that the unitary quantum system that we have constructed above is uniquely determined by the quasi-Hermitian operator $H$ and the metric operator $\etap$.

Next, we consider the restriction of the operator $\rho$ onto $\cS$. This gives a one-to-one linear operator that maps $\cS$ into $\sH$. Because $\cS$ is dense in $\sH_{\etap}$, we can view this operator as a densely-defined operator $\rho\big|_{_\cS}:\sH_{\etap}\to\sH$ having $\cS$ as its domain. In view of (\ref{inn-pro}), this is a bounded operator that can be extended to $\sH_\etap$ by continuity.\footnote{This is done by defining $\rho(\xi)$ for each $\xi\in\cH_\etap$ by taking a sequence $\{\xi_k\}$ in $\cS$ that converges to $\xi$ and identifying $\rho(\xi)$ with the limit of the sequence $\{\rho(\xi_n)\}$.} According to (\ref{inn-pro}), this bounded extension of $\rho\big|_{_\cS}$ that we denote by $\tilde\rho:\sH_\etap\to\sH$ is an isometry \cite[p 257]{kato}.

Let us use $\sR$ to label the range of $\tilde\rho$ which is a vector subspace of $\sH$. It is easy to show that $\sR$ is actually a closed subspace of $\sH$. To see this, we take a sequence $\{\xi_n\}$ in $\sR$ that converges to some $\xi\in\sH$. Clearly this is a Cauchy sequence, i.e., $\lim_{m,n\to\infty}\parallel\xi_m-\xi_n\parallel=0$. Now, let $\zeta_n:=\tilde\rho^{-1}\xi_n$. Then because $\tilde\rho$ is an isometry, we have
    \[\lim_{m,n\to\infty}\parallel\zeta_m-\zeta_n\parallel=
        \lim_{m,n\to\infty}\parallel\tilde\rho(\zeta_m-\zeta_n)\parallel=
        \lim_{m,n\to\infty}\parallel\xi_m-\xi_n\parallel=0.\]
Therefore $\{\zeta_n\}$ is a Cauchy sequence in $\sH_{_\etap}$. Because $\sH_{_\etap}$ is a Hilbert space, this sequence must converge to some $\zeta\in\sH_{_\etap}$. Now, recall that $\tilde\rho$ is bounded (continuous) linear operator. This implies that the sequence $\{\xi_n\}=\{\tilde\rho\,\zeta_n\}$ must converge to $\tilde\rho\,\zeta$, and as a result $\xi=\tilde\rho\,\zeta\in\sR$. This completes the proof that $\sR$ is a closed subspace of $\sH$. The following are important consequences of this fact.
    \begin{itemize}
    \item[1)] $(\sR,\br\cdot|\cdot\kt)$ is a separable Hilbert space that we label by $\hat\sH$. In general this is a Hilbert subspace of $\sH$. It coincides with $\sH$ provided that the linear span of $\rho\,\psi_n$ is dense in $\sH$.
    \item[2)] The operator $\hat\rho:\sH_{_\etap}\to\sR$ defined by $\hat\rho(\psi):=\tilde\rho(\psi)$ is a unitary operator.
    \item[3)] $\{\hat\rho\,\psi_n|n\in\N\}$, which is the same as $\{\rho\psi_n|n\in\N\}$, is an orthonormal basis of $\hat\sH$.
    \end{itemize}

We can use $\hat\rho$ and $\hat H$ to define a self-adjoint operator acting in $\hat\sH$ that is related to $\hat H$ via a similarity transformation. This is the operator  $h:=\hat\rho \hat H \hat\rho^{-1}:\hat\sH\to\hat\sH$.
By construction the pairs $(\sH_{_\etap},\hat H)$ and $(\hat\sH,h)$ are unitary-equivalent, therefore they represent the same quantum system. Following \cite{jpa-2003,jpa-2004}, we therefore refer to $h$ as the equivalent Hermitian Hamiltonian to the quasi-Hermitian operator $H$, and call $(\sH_{_\etap},\hat H)$ and $(\hat\sH,h)$ the pseudo-Hermitian and Hermitian representations of the quantum system in question, respectively.\footnote{The elements of $\sH\setminus\sR$ do not enter the formulation of the quantum system defined by the pair $(\hat\sH,h)$. They may be viewed as representing ``unobservable states,'' because they do not belong to the domain of the observables.}

Next, we examine the application of our constructions for a very simple and well-known toy model with $\sH:=L^2(\R)$.

Let $\alpha\in\R$, $V$ be a real and even confining potential, $p:=-i\frac{d}{dx}$, and
            \be
            H:=\frac{1}{2}(p-i\alpha)^2+V(x).
            \label{z-H=}
            \ee
This is one of the oldest examples of non-Hermitian $\cP\cT$-symmetric Hamiltonians that have a real spectrum. It was initially introduced for modeling certain localization effects in condensed matter physics \cite{Hatano}, and is one of the earliest examples considered in the framework of pseudo-Hermitian quantum mechanics \cite{Ahmed}. For definiteness we will confine our attention to the exactly solvable case where $V(x):=\omega^2x^2/2$ and $\omega\in\R^+$. Then $H$ is $\etap$-pseudo-Hermitian for $\etap:=e^{2\alpha x}$. Both $\etap$ and its positive square root, $\rho:=e^{\alpha x}$, are clearly unbounded positive-definite operators. It is easy to show that the following are eigenvectors of $H$.
        \be
        \psi_n(x):=N_nH_n(\sqrt\omega\,x)e^{-\frac{\omega x^2}{2}-\alpha x},
        \label{z-eg-ve}
        \ee
where $n\in\N$, $N_n$ are normalization constants, and $H_n$ are Hermite  polynomials. We also note that
        \bea
         (\etap\psi_n)(x)&=&N_n H_n(\sqrt\omega\,x)e^{-\frac{\omega x^2}{2}+\alpha x},
        \label{z-q0}\\
        (\rho\,\psi_n)(x)&=&N_n H_n(\sqrt\omega\,x)e^{-\frac{\omega x^2}{2}}.
        \label{z-q1}
        \eea
Because $\etap\psi_n$ are square-integrable functions, $\psi_n$ belong to the domain of $\etap$, and our constructions apply.

For this model, the Hilbert space $\sH_{_\etap}$ is defined by Cauchy completing the inner-product space obtained by endowing the linear span of $\psi_n$ with the inner product:
    \be
    \bbr\phi,\psi\kkt:=\int_{-\infty}^\infty e^{2\alpha x}\phi(x)^*
    \psi(x)dx.
    \ee
According to (\ref{z-q1}), $\{\rho\psi_n|n\in\N\}$ is an orthonormal basis of $\sH$. This implies that the Hilbert space $\hat\sH$ coincides with $\sH$, and $\hat\rho:\sH_{_\etap}\to\sH$ is a unitary operator. We can also easily show that in this case $h=\frac{1}{2}(p^2+\omega^2x^2)$. Therefore, $(\sH_{_\etap},\hat H)$ is a pseudo-Hermitian representation of the simple harmonic oscillator that we usually represent by $(\sH,h)$.

In conclusion, in this article, we have offered a mathematically rigorous construction of the physical Hilbert space, the observables, and the equivalent Hermitian Hamiltonian for a pseudo-Hermitian quantum system defined by an unbounded metric operator. This construction that applies for quasi-Hermitian Hamiltonian operators $H$ with a discrete spectrum relies on the natural assumption that the eigenvectors of $H$ should belong to the domain of the metric operator. It generalizes the well-known constructions given originally in \cite{jpa-2004} for bounded metric operators and differs from the latter in the sense that whenever the metric operator is unbounded the physical Hilbert space is generally different from the reference Hilbert space not only as inner-product spaces but also as vector spaces and sets. This however does not cause any difficulty. On the contrary, as the above simple example shows, the results reported in this article show that most of the unjustified and careless treatments of unbounded metric operators that are carried out in the literature on this subject can be put on solid grounds. One may try to extend the constructions given in this paper to indefinite metric operators. This would lead to indefinite-metric quantum theories with an unbounded metric operator whose study requires a separate investigation of its own.

\vspace{.3cm}
\noindent\textbf{\emph{Remark:}} After the submission of this article for publication, I was informed of Ref.~\cite{fz} where the authors also consider unbounded metric operators. They postulate the existence of an equivalent Hermitian operator that in our notation corresponds to $\rho H \rho^{-1}$, where $\rho:=\sqrt\etap$ acts in the original reference Hilbert space $\sH$. They further demand that $\rho H \rho^{-1}$ has a real discrete spectrum and a set of eigenvectors that form an orthonormal basis of $\sH$. They call an operator $H$ with the above property ``well-behaved'' with respect to $\etap$. In general, for a given densely-defined closed linear operator $H$ that is  $\etap$-pseudo-Hermitian and has a real and discrete spectrum, $\rho H \rho^{-1}$ does not satisfy all of these properties. In fact, it may have a domain that is not even dense in $\sH$. This raises the problem of characterizing these so-called well-behaved operators, an important problem that is not addressed in \cite{fz}. By restricting to these well-behaved operators the authors of \cite{fz} essentially circumvent the real mathematical problems that one must face while dealing with unbounded metric operators. In particular they overlook the need for considering Hilbert spaces that are different from $\sH$ even as sets.

\vspace{.3cm}
\noindent {\em \textbf{Acknowledgments:}} I wish to thank Carl Bender for reminding me of the need for addressing the unbounded metric operators, particularly during my talk in the PT-symmetric Quantum Mechanics Symposium held in Heidelberg, 25-28 September 2011. I would like to express my gratitude to organizers of this symposium, particularly Maarten DeKieviet, for their hospitality. I am grateful to Ali \"Ulger and Gusein Guseinov for their most illuminating comments and suggestions. This work has been supported by the Turkish Academy of Sciences (T\"UBA).

\np

\end{document}